\documentclass[]{spie}  
\pdfoutput=1

\usepackage{tikz}
\usepackage{amsmath,amsfonts,amssymb}
\usepackage{graphicx}
\usepackage{physics}
\usepackage{quantikz}
\usetikzlibrary{quantikz2}
\usepackage{tikz}
\usepackage[colorlinks=true, allcolors=blue]{hyperref}
\usepackage{cleveref}
\usepackage{subcaption}
\usepackage{float}
\usepackage{pifont}

\title{Quantum Algorithms for Drone Mission Planning}

\author[1]{Ethan Davies}
\author[1]{Pranav Kalidindi}
\affil[1]{Thales UK, Reading, United Kingdom}

\authorinfo{Ethan Davies : E-mail: Ethan.Davies@uk.thalesgroup.com\\  Pranav Kalidindi: E-mail: PranavKalidindi@hotmail.com}

\begin{document} 
\maketitle

\begin{abstract}
Mission planning often involves optimising the use of ISR (Intelligence, Surveillance and Reconnaissance) assets in order to achieve a set of mission objectives within allowed parameters subject to constraints. The missions of interest here, involve routing multiple UAVs visiting multiple targets, utilising sensors to capture data relating to each target. Finding such solutions is often an NP-Hard problem and cannot be solved efficiently on classical computers. Furthermore, during the mission new constraints and objectives may arise, requiring a new solution to be computed within a short time period. To achieve this we investigate near term quantum algorithms that have the potential to offer speed-ups against current classical methods. We demonstrate how a large family of these problems can be formulated as a Mixed Integer Linear Program (MILP) and then converted to a Quadratic Unconstrained Binary Optimisation (QUBO). The formulation provided is versatile and can be adapted for many different constraints with clear qubit scaling provided. We discuss the results of solving the QUBO formulation using commercial quantum annealers and compare the solutions to current edge classical solvers. We also analyse the results from solving the QUBO using Quantum Approximate Optimisation Algorithms (QAOA) and discuss their results. Finally, we also provide efficient methods to encode to the problem into the Variational Quantum Eigensolver (VQE) formalism, where we have tailored the ansatz to the problem making efficient use of the qubits available.  
\end{abstract}

\keywords{Quantum Annealing, QAOA, VQE,  QUBO, Mission Planning}


\section{Introduction}

Techniques for efficiently generating routes for Unmanned Ariel Vehicles (UAV's) have many practical implications including delivering goods and image sensing.
Previous models \cite{toth2002vehicle, braekers2016vehicle}
 for similar problems focus on the logistics for vehicles on land, so do not directly translate into this setting as the vehicles are constrained to roads and other land related conditions, whereas UAV's motions have far fewer limitations. The mission planning problems we wish to solve are often variants of the many-agent travelling salesman problems \cite{yang2019solving,baykasouglu2016multi} with additional constraints \cite{campbell2023multi, corberan2024multidepot, poikonen2020multi,wang2019vehicle,poikonen2020mothership} which often then become NP-hard problems. Efficient classical solutions therefore seem unlikely, however there could be potential for speedups using quantum algorithms.
UAVs will be equipped with different sensors such as Electro-Optic (EO), Infrared (IR) and SAR (Synthetic Aperture Radar) sensors depending on the mission. Each sensor type has different characteristics in terms of range, resolution (in range and bearing) and all-weather capability etc. The UAVs will need to arrive within a given distance in order to capture sensor data and to return to a base station in order to download the captured data. 

\noindent By finding the optimal use of the resources, we allow the improvement in human machine teaming by alleviating the analyst of more nugatory tasks and thereby moving the analyst higher up the value chain. This in turn can yield reduced operator workload, improved situational awareness and ultimately greater mission efficacy.
To avoid committing to any specific drone mission problem, we formalise the problem as generally as possible, in order to encapsulate as many drone mission problems as possible.
For the use cases we imagine, drone mission planning consists of having several drones capable of performing many different operations, starting from several bases. Each drone has its own set of parameters such as battery life and functionality (possibly many). The goal will be to find a set of routes for all drones, such that they don't crash, all sites requiring a drone of specific type of drone will be visited by a drone capable of the required functionality, within some time window. All drones must then return to their required base before they run out of battery. We can allow for drones to return to base early so that they can recharge their batteries, allowing for multiple routes to performed by a single drone. The goal is to then minimize certain aspects of the routes, which can be the total distance covered, the number of drones used, or the time taken for all drones to permanently return to base.

\noindent This paper is divided into three main sections. Section \ref{sec:background} discusses Mixed Integer Linear Programming (MILP)\cite{benichou1971experiments, floudas2005mixed,vielma2015mixed }, Quadratic Unconstrained Binary Optimisation (QUBO)\cite{lewis2017quadratic} and how to convert a MILP model into a QUBO. In Section \ref{sec:milp} we provide a Mixed Integer Linear Program (MILP) formulation of the problem that corresponds to the most general cases of drone routing mission problems. We then demonstrate that our problem formulation can be solved on real quantum hardware on a small scale toy model. We then go on to demonstrate the larger, more realistic size problems that can be solved on current cutting edge classical solvers. In Section \ref{sec:fault_tolerant}, we instead focus on the potential for fault tolerant quantum algorithms to assist with solving our problem, specifically Variational Quantum Eigensolver (VQE)\cite{tilly2022variational,cerezo2022variational} and Quantum Approximate Optimisation Algorithms (QAOA)\cite{farhi2014quantum, zhou2020quantum,blekos2024review} inspired quantum circuits to solve the standard Travelling Salesman Problem (TSP) in which we simulate the algorithms for much smaller problems and comment on  their effectiveness and potential to be used on larger quantum devices.

\section{Background}
\label{sec:background}


In this paper we investigate the potential of quantum solvers for the travelling salesman problems and their variants\cite{qian2023comparative}. This is an inherently classical problem and we will need to use a series of reductions from the original problem formulation in order to be compatible with quantum algorithms. Here, we outline the several problems and explain how they can be converted into each other.

\subsection{Mixed Integer Linear Program (MILP)}

A Mixed Integer Linear Program (MILP) aims to minimize the cost function $c^Tx$, subject to linear constraints that a feasible solution $x$ must satisfy.

\begin{equation}
    \begin{split}
    \min_{x\in \mathbb{Z}^n} &\quad c^{T}x \\
    \text{such that} &\quad l \leq Ax \leq u,\quad x_l \leq x \leq x_u, \\
    \text{where} &\quad x \in \mathbb{Z}^n,\, A\in \mathbb{Z} ^{n\times m},\,\, l,u \in \mathbb{Z}^m
    \label{eq:milp_eq}
    \end{split}
\end{equation}
Some of the best known solvers of MILP problems are Gurobi\cite{jablonsky2015benchmarks} and CPLEX \cite{cplex-ibm}, which are commercial solutions. These utilise the branch and cut algorithm to efficiently find solutions to the MILP problems. Gurobi performs much better on average \cite{sun2024mindoptadaptercplexbenchmarking} compared to open source solutions such as GLPK \cite{glpk}. Gurobi is used to benchmark our quantum algorithms.
Despite the idea that linear constraints can seem restrictive to what problems can be cast into this form, by increasing the dimension of $x$, non-linear constraints can become linearized.

\begin{itemize}
\item Suppose we have a  binary bit $b\in \{0,1\}$ and we wish $c^Tx \leq d$ to hold in the case that $b=1$. We achieve this by picking a large enough constant $M$ such that $\forall x,\quad  c^Tx - M \leq d$. We can then use the inequality $c^Tx - (1-b)M \leq d$, to achieve the required result.
\item Suppose we have two inequalities $c_0^T x \leq d_0$ and $c_1^T x \leq d_1$, and we wish for at least one of these inequalities to hold but have no preference. We introduce an extra binary variable $b\in \{0,1\}$ and use it to turn off one of the equalities. We would then instead use $c_0^T x - (1-b)M_0 \leq d_0$ and $c_1^T x -b M_1 \leq d_1$ instead. 
\item Suppose we have a binary bit $b\in\{0,1\}$ and we want the equality $c^Tx = d$ to hold in the event $b=1$. We can achieve this by asking for the weak inequality to hold in both directions when $b=1$. Here we will use  $c^Tx - (1-b)M_0 \leq d$ and $c^Tx + (1-b)M_1 \geq d$
\item Suppose we have several bounded integer variables $X_i$ and we wish to find $\min \max_i X_i$. We can achieve this by introducing an additional variable $X$ and then solve the problem $\min X \, s.t.\,\forall i\,  X_i \leq X $
\item If we have an MILP defined for some larger problem and then obtain some new information fixing some of our variables to either exact integers or a tighter bound, the resulting problem will still be a MILP formulation over a smaller space.
\item If we want to obtain higher order expressions than linear, it can be achieved. Extra variables would be introduced and constrained to be the product of other terms.
The case we will be using is $z = x \times y,\, x,y,z \in \{0,1\}$. This can be achieved with the following constraints;
\begin{itemize}
    \item[$\bullet$] $1+z \geq x+y $
    \item[$\bullet$] $z\leq x$
    \item[$\bullet$] $z\leq y$
\end{itemize}
\end{itemize}

\subsection{Quadratic Unconstrained Binary Optimisation (QUBO)}

Quadratic Unconstrained Binary Optimisation (QUBO) problems aim to minimize the value of $x^{T}Qx$ where $x\in \mathbb{Z}_2^n$ is a binary vector and $Q\in\mathbb{R}^{n\times n}$  is a real valued symmetric matrix. With a polynomial overhead in size, a MILP instance can be converted into a QUBO\cite{davies2024optical}. This is achieved by using binary expansions of all integer variables. Each linear constraint, $C_i$ is then manipulated into their own QUBO, $P_i$, which is minimised exactly when the constraint is satisfied. The resulting QUBO from the MILP formulation is then given by $Q = \text{diag}(c) +\sum_i \lambda_i P_i$, where $\lambda_i$ are hyperparameters that weight all the constraints, these must be chosen high enough, such that it is never beneficial to violate a constraint in order to reduce $c^T x$.
There exist many classical heuristic algorithms for QUBOs such as Gurobi, and simulated annealing\cite{heim2015quantum}.

\subsection{Ground State Problem}
\label{quantum algorithms}

The Ground State Problem (GSP) \cite{li2011solving, szulkin2009ground} is given a Hamiltonian, $H$, find the quantum state $\ket{\psi}$, which minimizes $\bra{\psi} H \ket{\psi}$. A QUBO instance can be converted into this problem by using a change of variables $x_i \rightarrow z_i = 1-2x_i$. The function we wish to minimize is then $x^TQx = \sum_{i,j} Q_{i,j}x_ix_j = \sum_{i,j} \tilde{Q}_{i,j}z_i z_j$, where $\tilde{Q}$ is the transformation of $Q$. We can write the QUBO instance as the Ground State Problem with $H = \sum_{i,j} \tilde{Q}_{i,j}Z_{i}Z_{j}$.
When benchmarking these problems, the quantity used for comparisons is the Average Ratio (AR), $AR(\ket{\psi}) = \frac{\bra{\psi}H\ket{\psi}}{H_{min}}$. It is hoped that if we obtain state $\ket{\psi}$ with AR close to 1, measuring this state will then yield close to optimal solutions to the corresponding QUBO.
There are many different approaches suggested for this problem including:

\begin{itemize}
    \item Quantum annealing\cite{rajak2023quantum,hauke2020perspectives}. This exists in a weaker paradigm of quantum computing that does not yield universal computation. A quantum state in the ground state of $H_X$ is constructed and evolved under the Hamiltonian $H(t) = (1-f(t))H_X + f(t)H_G$, where $f(0)=0, f(T)=1$. As long as $f$ varies sufficiently slowly, the state $\ket{\psi(t)}$ should remain close to the ground state of $H(t)$ due to the adiabatic invariance\cite{lenard1959adiabatic}. The goal is to encode a solution to the ground state of the $H_G= \sum_i \alpha_i Z^z_i$. This way the ground state is guaranteed to be in the computational basis and and can be obtained directly by measuring $\ket{\psi(t)}$. Current technology limits the weight of $z_i$ to be at most 2, so the problem we are interested in must be formulated as a Quadratic Unconstrained Binary Optimisation (QUBO).

    We evolve the state for time,  $T$, and then measure the state $\ket{\psi(T)}$ in the computational basis and read off the ground state.

    \item Variational Quantum Eigensolver. This approach is to use a parameterised quantum circuit $U(\vec{\theta})$ to construct the quantum state $\ket{\psi(\vec{\theta})} = U(\vec{\theta}) \ket{\psi_0}$. Expectations are then taken to approximate $\bra{\psi(\vec{\theta})} H \ket{\psi(\vec{\theta})}$. Classical machine learning approaches are then chosen to update $\vec{\theta}$ in order to minimize $\langle H \rangle $.
    \item
    
    Quantum Approximate Optimisation Algorithm (QAOA), which is a special type of VQE where the ansatz circuit is chosen to mimic the Lie-Trotter factorisation of $e^{i(A(t)H_{fin}  + B(t)H_{mixer})}$, so in principle can simulate quantum annealing. The circuit has parameters $\vec{\gamma}, \vec{\beta}$ and the circuit is alternating blocks of $e^{i\gamma_i H_{fin}}$ and $e^{i\beta_i H_{mixer}}$. Expectations of the final state are then computed and used with a  classical optimiser to train $\vec{\gamma} , \vec{\beta}$.

\end{itemize}

\section{Annealing Based Approach}
\label{sec:milp}

\subsection{Problem Definition}
\label{sec:problem_def}
We mathematically model the problem as follows. Let $G$ be a complete, undirected, multi-dimensional weighted graph, where the nodes, $V$, consist of bases and locations to visit, and edges $E$ be the possible paths between nodes (travelling in straight line) with associated edge parameters, such as the time to travel between vertices, the battery used to traverse the distance, or cost of taking this edge. Let $K$ be a set of labels of functionalities (representing the diverse set of functions the group of drones can perform). Each drone, $\text{drone}_i$, has a base vertex, $v_i\in V$, in which the drone must begin and end its journey.
A valid route set is a set $\Gamma$ of routes in $G$ such that for every drone $d_i \in D$, there exists a route  $\gamma \in \Gamma$, starting and finishing at $b_i$ such that the drone has not exceeded its battery or capabilities. The drone has the correct functionalities for vertices that it visits and performs the tasks at. The total time taken for the journey is below $T_{max}$. The goal is to then to find the optimal route set $\Gamma$ such that it yields the minimum possible $T_{max}$.

\subsection{Formulation of MILP Model}

In this section we use the following notation for the problem defined in \cref{sec:problem_def}:

We model the problem as a complete graph, where the nodes are locations that the drones may visit. Certain edges and nodes belong to groups that possess different characteristics.

\begin{itemize}
    \item $V_O$ - These are the nodes with objectives that must be visited during the mission.
    \item $V_I$ - These are the nodes which are intermediate, they don't have to be visited. This includes nodes at which drones may recharge but also be additional nodes added to the problem in an attempt to produce smoother routes for the drones.
    \item $V_S$ - These are nodes at which a drone starts the mission.
    \item $V_E$ - These are nodes at which a drone must end the mission.
    \item $V_{rec}$ - These are nodes at which drones can recharge their batteries.
    \item $E_O$ - These are edges with objectives that must be traversed (in either direction).
    \item $E_I$ - These are edges which are intermediate. They don't have to be traversed.
\end{itemize}

The important properties of the graph are:
\begin{itemize}
    \item $T_{a,b}$ - The time taken to traverse from $a$ to $b$ and then complete the objective at $b$.
    \item $B_{a,b}$ - The battery used to traverse from $a$ to $b$ and then complete the objective at $b$.
    \item $q_{a}$ - The quantity of the goods the drone must use for the objective at $a$. This could be memory or payload.
\end{itemize}

Each drone also has its own set of parameters:
\begin{itemize}
    \item $B_{max}$ - The maximum amount of battery the drone can possess. 
    \item $B_{hov}$ - The rate at which battery is used whilst hovering.
    \item $Q_{max}$ - The maximum quanity of goods the drone can possess.
    \item $B_{recharge}$ - The rate at which the drone recharges.
    \item $D$ - A  possibly unique ID that contains binary variables of what capabilities the drone possesses.

\end{itemize}

The goal is to obtain a family of routes that the drones can take, such that each objective is met by a drone capable of fulfilling the mission. We then wish to minimize the time taken for all drones to return to their respective bases.
Previous MILP formulations for similar problems\cite{campbell2023multi} create a set of variables for each drone route independently and then constrain them against each other to ensure that all objectives are met. This would create a solution that has $O(DE)$ variables. The alternative option is to keep track of the drones journeys by storing information at each node. Constraining nodes connected by an edge to be consistent with their information will then allow us to construct valid routes. This approach will create $O(E)$ variables which could create smaller formulations of the desired mission when the number of drones becomes too large.
We make the choice of reformulating the graph as follows. For each drone in each base node, we replace the base node with a start node, $v_s$, an end node $v_e$ and multiple recharge nodes $v_{rec}$. Properties such as time and battery usage will be inherited from the original base node.
We can now define our integer variables.

\begin{itemize}
    \item $x_{a} \in \{0,1\}$ is 1 if node $a$ is visited by any of the drones
    \item $e_{(a,b)} \in \{0,1\}$ is 1 if edge $(a,b)$ is used by any of the drones
    \item $T_{a}\in [0,T_{max}]$ is the time at which the drone leaves node $a$
    \item $B_{a}\in [0,B_{max}]$  serves as a lower bound on the remaining battery when the drone reaches $a$ 
    \item $Q_{a}\in [0, Q_{max}]$ serves as a lower bound on the remaining resource of some quantity when the drone reaches node $a$
    \item $D_{a}\in \{0,1\}^r$ is the binary string corresponding to the node that visited node $a$, The first $k$ bits of $D_{a}$ correspond to whether the drone at node $a$ has capability $K_i$ in the $i^{\text{th}}$ bit. The remaining bits are then the unique ID of the specific drone, if we want to ensure a specific drone returns back to a specific base.
\end{itemize}

With these variables in hand, we can now give the constraints for a valid set of drone routes:

\begin{itemize}
    \item All objective nodes are reached:
    \begin{itemize}
        \item $\forall a \in V_O \setminus V_E\quad \sum_{b}e_{(a,b)} =1$
        \item $\forall a \in V_O \setminus V_S\quad \sum_{b}e_{(b,a)}=1$
    \end{itemize}
    \item All intermediate nodes are consistent - if a node is entered then it must be left:
    \begin{itemize}
        \item $\forall a \in V_I\quad \sum_{b} e_{(a,b)}=x_a$
        \item $\forall a \in V_I \quad  \sum_{b} e_{(b,a)}=x_a$
    \end{itemize}
    \item Times at nodes are consistent:\
    \begin{itemize}
        \item $\forall (a,b) \in E\quad  t_b \geq t_a +T_{(a,b)} - (1-e_{(a,b)})M_T$.
    \end{itemize}
    Here, time consistency prevents sub-tours from forming. A drone starting in the middle of the graph and completing a loop to itself would satisfy the previously mentioned edge conditions, the time constraints enforce a total ordering on the nodes, which is increasing along any path taken, thus forbidding loops from forming.
    It is also worth noting here that despite the constraint being `turned on', the inequality not being tight can have additional meaning. The additional time $\delta_{(a,b)} = t_b - t_a - T_{(a,b)}\geq 0$ is interpreted as additional time that the drone has chosen to wait at $a$ before heading off on the next node in the journey. It may choose to do this for two main reasons, to avoid a collision with another drone or to charge back up if $a$ is a charging point for the drone.

    \item Battery values are consistent:\\
    There are two cases for battery consistency we must deal with. Along normal edges, the battery must decrease by at least $B_{(a,b)}$. The drone also has the potential to hover at the node. Unlike the other two terms, this will be proportional to the additional time spent at node $b$, perhaps to prevent a collision.
    \begin{itemize}
        \item $\forall (a,b) \in E,\,  a\notin V_R,\quad  B_b \leq B_a -B_{(a,b)} - B_{hover}\delta_{(a,b)} + (1-e_{(a,b)})M_B$
    \end{itemize}
    
    The second case is that drone is leaving a charging station. Now the battery has the potential to recharge. 
    \begin{itemize}
        \item $B_b \leq B_a + \delta_{(a,b)}B_{recharge}  -  B_{(a,b)} + (1-e_{(a,b)})M_B $
        \item $B_a + \delta_{(a,b)}B_{recharge}  \leq B_{max} + (1-e_{a,b})M_B$
    \end{itemize}
    \item Resource usage is consistent:
    \begin{itemize}
        \item $\forall (a,b) \in E,\,  b\notin V_R,\quad Q_b  - q_{a} < Q_a + (1-e_{(a,b)})M_Q$
    \end{itemize}
    In the case $b\in V_r$, we have no constraint and the quantity value is free to be anything when leaving the recharge node. This is because the data can be downloaded at this stage, or the payload can be renewed.
    \item Drone IDs are consistent  between nodes:
    \begin{itemize}
        \item $\forall (a,b) \in E\quad D_b \geq D_a +(1-e_{(a,b)})M_d$
         \item $\forall (a,b) \in E\quad D_b \leq D_a -(1-e_{(a,b)})M_d$
    \end{itemize}

    \item Objectives with specific needs are met by drones with the correct functionalities. This can be achieved by partially fixing the drone ID associated with the node. When an edge requires specific needs, we partially fix the ID with both nodes on either end. The entire ID of the node can be fixed at start, recharge and end points in order to ensure the correct drones are starting and finishing in the correct base.
    
    \item Drones do not crash into each other. If we can set the altitudes of all drones to be different, this constraint would not be needed. However, we can still deal with this constraint in our model by placing an additional constraint on pairs of edges that intersect. Suppose we are concerned about two drones crashing when travelling down $(a,b)$ and $(c,d)$. We will stop crashing from happening by insisting that in the case both edges are traversed, either the first drone has reached $b$ before the second drone leaves $c$ or the second drone has reached $d$ before the first drone leaves $a$. 
    We achieve this with the following equations.
    \begin{itemize}
        \item $y_{(a,b),(c,d)} = e_{(a,b)}\times e_{(c,d)}$
        \item $t_c \geq t_a + T_{(a,b)} - (1-y_{(a,b),(c,d)})M - b_{(a,b),(c,d)}M$
        \item $t_a \geq t_c + T_{(c,d)} - (1-y_{(a,b),(c,d)})M - (1-b_{(a,b),(c,d)})M$
    \end{itemize}

\item Time windows can be included. If a node, $a$, must be visited within a certain time frame, $(t_{\min},t_{\max} )$. We restrict $t_a \in [t_{\min}, t_{\max}]$.

\item The goal is often to minimize the total time for all drones to return to their respective bases, we introduce an additional variable $T$ and include the constraint $\forall a \in V_E\quad T_a \leq T$. The objective function we then wish to minimize is then $T$.
\item If the goal is to minimize the total energy (battery power) spent on the mission, the objective function to minimize would then be $\sum_{(a,b)\in E} B_{(a,b)}e_{(a,b)}$.
\end{itemize}

\subsection{Toy Example with Quantum Annealing}

\begin{figure}[h]
\begin{subfigure}{0.5\textwidth}
    \begin{tikzpicture}[circle,
    ->, 
    >=Stealth, 
    every node/.style={draw,minimum size=0.2cm, font=\sffamily\Large\bfseries} 
,scale = 3]
\small 
\node[] (A) at (0, 1) {A};
\node[] (B) at (-1,0) {B};
\node[] (C) at (1,0) {C};
\node[] (D) at (0,-1) {D};

\draw (A) -- (B) node[near start,left , draw=none] {$(2,1)$};
\draw (B) -- (A);
\draw (B) -- (C);
\draw (C) -- (B)node[near end , yshift = 0.4cm, draw=none] {$(6,2)$};
\draw (C) -- (D);
\draw (D) -- (C)node[near start,right, draw=none] {$(1,2)$};;
\draw (D) -- (A);
\draw (A) -- (D) node[near start , xshift = 0.6cm, draw=none] {$(2,2)$};;
\draw (A) -- (C) node[near start, right, draw=none] {$(3,1)$};
\draw (C) -- (A);
\draw (D) -- (B) node[near start,left, draw=none] {$(5,1)$};
\draw (B) -- (D);
\end{tikzpicture}
\caption{}
\label{fig:subim1}
\end{subfigure}
\begin{subfigure}{0.5\textwidth}

    \begin{tikzpicture}[circle,
    ->, 
    >=Stealth, 
    every node/.style={draw,minimum size=0.2cm, font=\sffamily\Large\bfseries} 
,scale = 3]
\small 
\node[] (A_start) at (0, 1) {A};
\node[minimum size=0.8cm] (A_rec) at (0, 1) {};
\node[minimum size=0.8cm] (A_end) at (0, 1) {};
\node[] (B) at (-1,0) {B};
\node[] (C) at (1,0) {C};
\node[] (D) at (0,-1) {D};

\node[above=0.cm of B, font=\small, draw = none] {$(2,3)$};
\node[left=0.cm of D, font=\small, draw = none] {$(7,2)$};
\node[left=0.cm of A_rec, font=\small, draw = none] {$(0,4)$};
\node[below right=0.cm of A_rec, xshift = -0.2cm, yshift = -0.3cm, font=\small, draw = none] {$(9,0)$};

\node[above =0.cm of A_rec,  yshift = -0.3cm ,font=\small, draw = none] {$(11,2)$};

\node[right =0.cm of C, font=\small, draw = none] {$(14,1)$};

\node[right =0.cm of A_end, yshift =-0.4cm,font=\small, draw = none] {$(17,0)$};

%

\draw (A_start) -- (B);
\draw (B) -- (D);
\draw (D) -- (A_rec);
\draw (A_rec) to [out=0,in=90] (C);
\draw[bend right] (C) -- (A_end);

\end{tikzpicture}

\end{subfigure}

\caption{(Left) Toy instance of drone mission with a single drone. Each edge has two associated parameters, the time taken to traverse the edge and the battery expended in the process. The drone in the problem has an initial battery of 4 units. (Right) One of the optimal routes the drone can take. At each node we show the current time and remaining battery level of the drone. }
\label{dwave toy}

\end{figure}
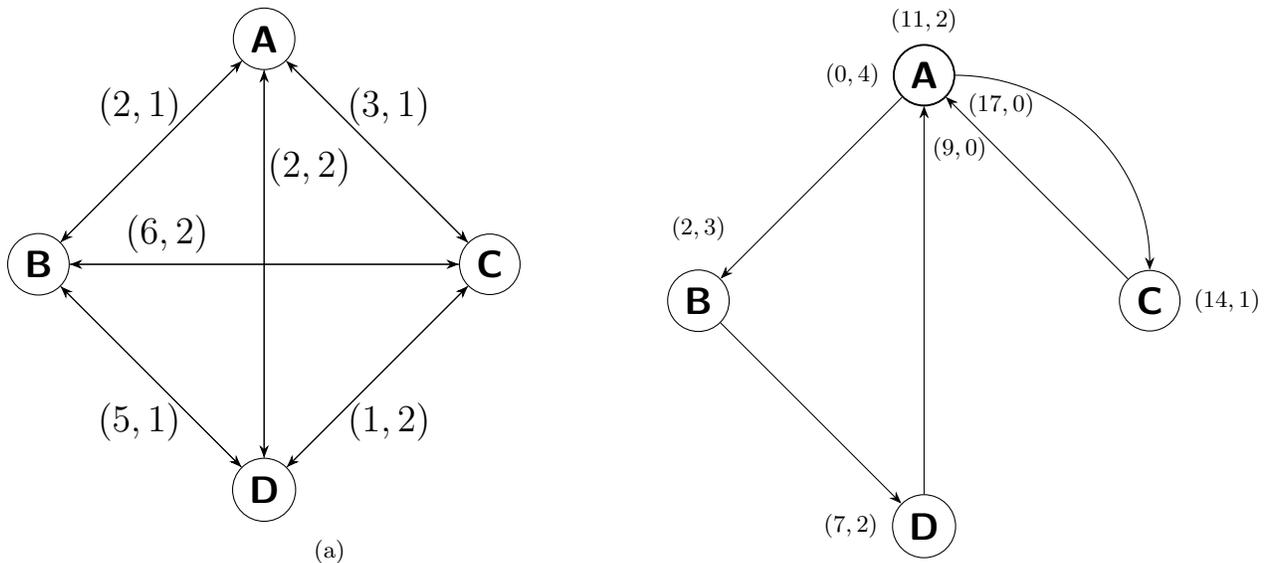

Problem instances that can be run on current quantum annealers are constrained by their size as well as their decoherence times and the connectivity between qubits. This makes any moderately sized instance of the problem too large to run on current hardware. Instead, we attempt to solve the toy model in \cref{dwave toy} on the D-Wave Leap's Hybrid solver. The solver combines quantum annealing with classical algorithms in a hybrid approach which is kept private by D-Wave, limiting what we conclusions we can speculate from the results. We encoded the problem into a MILP instance as in \cref{eq:milp_eq} and then converted into a QUBO with 275 variables for us to then submit to the D-Wave platform. Out of the 100 attempts to solve the problem with an average processing time of 3 seconds on the Leap's Hybrid solver, which is a combination of both quantum  and classical processing times, we were able to find a valid solution 4 times, and all of them were optimal solutions. More time can be spent tuning the hyperparameters of the QUBO to balance the constraints, but this is left for future work.

\subsection{Toy Example with Classical Solver}

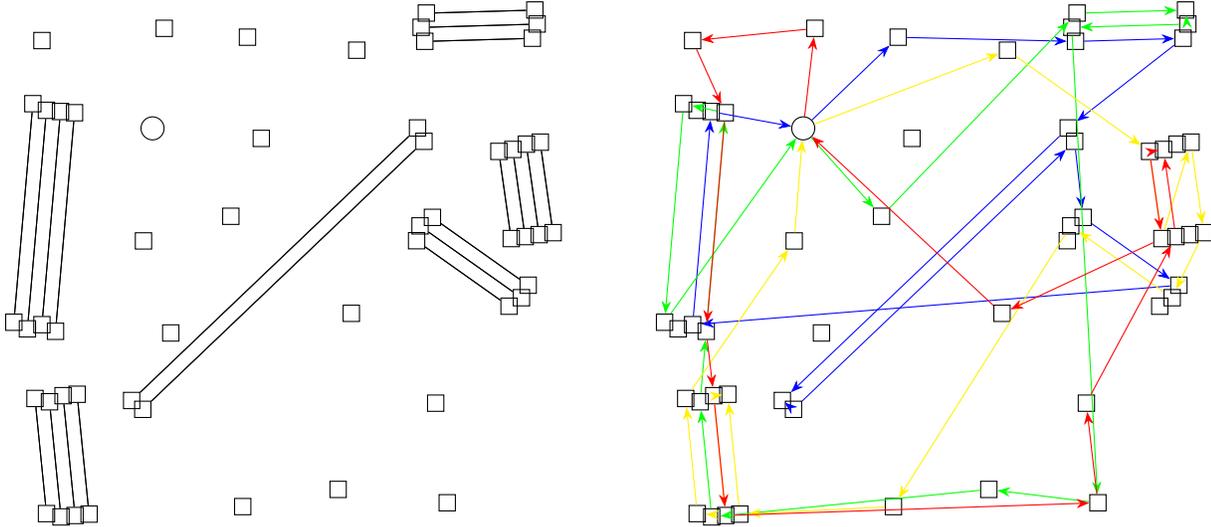
\begin{figure}[H]
\begin{subfigure}{0.5\textwidth}

\begin{tikzpicture}[
    -, 
    >=Stealth, 
    every node/.style={draw,minimum size=0.2cm, font=\sffamily\Large\bfseries} 
,scale = 0.0125]
\small 
\node[] (1) at (29.8848, 507.122) {};
\node[] (2) at (159.979, 520.148) {};
\node[] (3) at (248.523, 510.737) {};
\node[] (4) at (364.802, 496.87) {};
\node[] (5) at (433.247, 521.246) {};
\node[] (6) at (556.652, 524.775) {};
\node[] (7) at (49.6156, 431.582) {};
\node[circle] (8) at (147.536, 413.665) {};
\node[] (9) at (263.198, 402.976) {};
\node[] (10) at (436.281, 399.902) {};
\node[] (11) at (545.205, 396.87) {};
\node[] (12) at (137.903, 293.839) {};
\node[] (13) at (230.826, 320.063) {};
\node[] (14) at (432.098, 309.854) {};
\node[] (15) at (558.82, 300.537) {};
\node[] (16) at (29.8873, 204.664) {};
\node[] (17) at (166.929, 195.818) {};
\node[] (18) at (358.982, 216.733) {};
\node[] (19) at (539.982, 233.37) {};
\node[] (20) at (52.3489, 129.08) {};
\node[] (21) at (137.158, 114.523) {};
\node[] (22) at (448.742, 121.091) {};
\node[] (23) at (65.0995, 1.4916) {};
\node[] (24) at (243.48, 10.9768) {};
\node[] (25) at (344.956, 29.2744) {};
\node[] (26) at (461.023, 15.0345) {};
\node[] (27) at (437.176, 506.352) {};
\node[] (28) at (551.581, 509.624) {};
\node[] (29) at (438.818, 536.411) {};
\node[] (30) at (554.223, 539.711) {};
\node[] (31) at (34.672, 432.881) {};
\node[] (32) at (14.4436, 200.212) {};
\node[] (33) at (64.5593, 430.282) {};
\node[] (34) at (44.3309, 197.614) {};
\node[] (35) at (20.2284, 439.931) {};
\node[] (36) at (0.0, 207.263) {};
\node[] (37) at (429.427, 414.094) {};
\node[] (38) at (125.304, 123.945) {};
\node[] (39) at (530.853, 391.233) {};
\node[] (40) at (543.968, 298.438) {};
\node[] (41) at (560.058, 398.969) {};
\node[] (42) at (573.672, 302.636) {};
\node[] (43) at (516.0, 389.134) {};
\node[] (44) at (529.115, 296.339) {};
\node[] (45) at (428.423, 294.072) {};
\node[] (46) at (526.806, 224.323) {};
\node[] (47) at (445.273, 318.9) {};
\node[] (48) at (547.157, 246.67) {};
\node[] (49) at (37.9232, 122.586) {};
\node[] (50) at (50.1739, 0.0) {};
\node[] (51) at (67.2745, 130.572) {};
\node[] (52) at (80.0252, 2.9832) {};
\node[] (53) at (22.4976, 126.097) {};
\node[] (54) at (34.7482, 3.51163) {};
\draw (5) -- (6);
\draw (6) -- (5);
\draw (7) -- (16);
\draw (16) -- (7);
\draw (10) -- (21);
\draw (21) -- (10);
\draw (11) -- (15);
\draw (15) -- (11);
\draw (14) -- (19);
\draw (19) -- (14);
\draw (20) -- (23);
\draw (23) -- (20);
\draw (27) -- (28);
\draw (28) -- (27);
\draw (29) -- (30);
\draw (30) -- (29);
\draw (31) -- (32);
\draw (32) -- (31);
\draw (33) -- (34);
\draw (34) -- (33);
\draw (35) -- (36);
\draw (36) -- (35);
\draw (37) -- (38);
\draw (38) -- (37);
\draw (39) -- (40);
\draw (40) -- (39);
\draw (41) -- (42);
\draw (42) -- (41);
\draw (43) -- (44);
\draw (44) -- (43);
\draw (45) -- (46);
\draw (46) -- (45);
\draw (47) -- (48);
\draw (48) -- (47);
\draw (49) -- (50);
\draw (50) -- (49);
\draw (51) -- (52);
\draw (52) -- (51);
\draw (53) -- (54);
\draw (54) -- (53);
\end{tikzpicture}

\label{fig:subim1}
\end{subfigure}
\begin{subfigure}{0.5\textwidth}

\begin{tikzpicture}[
    ->, 
    >=Stealth, 
    every node/.style={draw,minimum size=0.2cm, font=\sffamily\Large\bfseries} 
,scale = 0.0125]
\small 
\node[] (1) at (29.8848, 507.122) {};
\node[] (2) at (159.979, 520.148) {};
\node[] (3) at (248.523, 510.737) {};
\node[] (4) at (364.802, 496.87) {};
\node[] (5) at (433.247, 521.246) {};
\node[] (6) at (556.652, 524.775) {};
\node[] (7) at (49.6156, 431.582) {};
\node[circle ] (8) at (147.536, 413.665) {};
\node[] (9) at (263.198, 402.976) {};
\node[] (10) at (436.281, 399.902) {};
\node[] (11) at (545.205, 396.87) {};
\node[] (12) at (137.903, 293.839) {};
\node[] (13) at (230.826, 320.063) {};
\node[] (14) at (432.098, 309.854) {};
\node[] (15) at (558.82, 300.537) {};
\node[] (16) at (29.8873, 204.664) {};
\node[] (17) at (166.929, 195.818) {};
\node[] (18) at (358.982, 216.733) {};
\node[] (19) at (539.982, 233.37) {};
\node[] (20) at (52.3489, 129.08) {};
\node[] (21) at (137.158, 114.523) {};
\node[] (22) at (448.742, 121.091) {};
\node[] (23) at (65.0995, 1.4916) {};
\node[] (24) at (243.48, 10.9768) {};
\node[] (25) at (344.956, 29.2744) {};
\node[] (26) at (461.023, 15.0345) {};
\node[] (27) at (437.176, 506.352) {};
\node[] (28) at (551.581, 509.624) {};
\node[] (29) at (438.818, 536.411) {};
\node[] (30) at (554.223, 539.711) {};
\node[] (31) at (34.672, 432.881) {};
\node[] (32) at (14.4436, 200.212) {};
\node[] (33) at (64.5593, 430.282) {};
\node[] (34) at (44.3309, 197.614) {};
\node[] (35) at (20.2284, 439.931) {};
\node[] (36) at (0.0, 207.263) {};
\node[] (37) at (429.427, 414.094) {};
\node[] (38) at (125.304, 123.945) {};
\node[] (39) at (530.853, 391.233) {};
\node[] (40) at (543.968, 298.438) {};
\node[] (41) at (560.058, 398.969) {};
\node[] (42) at (573.672, 302.636) {};
\node[] (43) at (516.0, 389.134) {};
\node[] (44) at (529.115, 296.339) {};
\node[] (45) at (428.423, 294.072) {};
\node[] (46) at (526.806, 224.323) {};
\node[] (47) at (445.273, 318.9) {};
\node[] (48) at (547.157, 246.67) {};
\node[] (49) at (37.9232, 122.586) {};
\node[] (50) at (50.1739, 0.0) {};
\node[] (51) at (67.2745, 130.572) {};
\node[] (52) at (80.0252, 2.9832) {};
\node[] (53) at (22.4976, 126.097) {};
\node[] (54) at (34.7482, 3.51163) {};
\draw[blue] (48) -- (16);
\draw[blue] (16) -- (7);
\draw[blue] (27) -- (28);
\draw[blue] (8) -- (3);
\draw[blue] (7) -- (8);
\draw[blue] (37) -- (38);
\draw[blue] (47) -- (48);
\draw[blue] (21) -- (10);
\draw[blue] (3) -- (27);
\draw[blue] (38) -- (21);
\draw[blue] (28) -- (37);
\draw[blue] (10) -- (47);
\draw[green] (36) -- (8);
\draw[green] (26) -- (25);
\draw[green] (33) -- (35);
\draw[green] (30) -- (6);
\draw[green] (5) -- (26);
\draw[green] (49) -- (34);
\draw[green] (50) -- (49);
\draw[green] (34) -- (33);
\draw[green] (6) -- (5);
\draw[green] (29) -- (30);
\draw[green] (35) -- (36);
\draw[green] (25) -- (50);
\draw[green] (8) -- (13);
\draw[green] (13) -- (29);
\draw[yellow] (24) -- (52);
\draw[yellow] (14) -- (24);
\draw[yellow] (23) -- (54);
\draw[yellow] (52) -- (51);
\draw[yellow] (42) -- (19);
\draw[yellow] (53) -- (12);
\draw[yellow] (43) -- (44);
\draw[yellow] (12) -- (8);
\draw[yellow] (51) -- (20);
\draw[yellow] (44) -- (41);
\draw[yellow] (8) -- (4);
\draw[yellow] (41) -- (42);
\draw[yellow] (19) -- (14);
\draw[yellow] (4) -- (43);
\draw[yellow] (20) -- (23);
\draw[yellow] (54) -- (53);
\draw[red] (34) -- (20);
\draw[red] (8) -- (2);
\draw[red] (22) -- (40);
\draw[red] (44) -- (18);
\draw[red] (26) -- (22);
\draw[red] (23) -- (26);
\draw[red] (2) -- (1);
\draw[red] (39) -- (43);
\draw[red] (20) -- (23);
\draw[red] (1) -- (33);
\draw[red] (40) -- (39);
\draw[red] (18) -- (8);
\draw[red] (33) -- (34);
\draw[red] (43) -- (44);
\end{tikzpicture}
\label{fig:subim2}
\end{subfigure}

\caption{(Left) Multi-agent TSP instance. Four drones must start and end at the base (circle). Edges shown must be traversed in either direction. (Right) Solution found using Gurobi with a running time of 180 seconds.}
\label{Gurobi-big}

\end{figure}

Current classical solvers do not possess the same size constraints as quantum solvers and can be run on problems of thousands of variables with thousands of constraints, so we are able to run the MILP formulation of the problem on much larger problem instances. We were able to find efficient solutions to the problem instance in \cref{Gurobi-big}. One potential issue we encountered was, since we were optimising over the time for the last drone to return to base, there was no incentive for the penultimate (or in fact any faster drone) to return to the base earlier than is required. Some drones would wait at intermediate nodes until all nodes can return at the same time. This could be resolved by not allowing the drone to wait to complete the mission once the paths are chosen, but this would not help with the faster drones taking inefficient paths.
To solve this we ran multiple problem instances in succession as follows.

\begin{enumerate}
    \item Run the solver on the full MILP.
    \item For each drone route found, run the solver on a reduced MILP (with only the one drone, that must visit the same objective nodes and edges, but can now take a more efficient route) as the MILP is smaller and the solver can be more effective.
    \item Take the individual drone route with the longest time to return to base and fix the route. Remove the required nodes and edges visited from full MILP formulation and reduce the number of drones by one.
    \item If there are any drones without a set route fixed, return to Step 1.
    \item If required, we run all of the routes together with the `no crashing' constraints to then ensure the drones wait at intermediate nodes to prevent crashes.
\end{enumerate}
Breaking the problem into many smaller problem instances to solve can be useful to guide the solver towards good solutions, as we understand the structure of the problem and how the constraints are interlaced, a lot more intuitively than algorithm can. Leaving the no `no crashing' constraints to the end can lead to sub-optimal solutions as there can be many intersections between routes and congestion in the drones routes. 
Choosing how to allocate the total processing time between intermediate instances is an important aspect of this algorithm and should be top heavy towards the larger instances of the problem.

\section{Quantum Algorithms for The  Travelling Salesman Problem}
\label{sec:fault_tolerant}
In this section, we focus solely on standard TSP and suggest ways in which we can modify existing quantum algorithms to respect the constraints of TSP.
\subsection{Hamiltonian Formulation}
For these algorithms, we focus on the time indexed formulation of travelling salesman problem \cite{lucas2014ising}.
The main point in this formulation is that a route can described using a permutation matrix $P_{\sigma} \in \mathbb{R}^{n\times n},\quad {P_{\sigma}}_{i,j} =\delta_{j, \sigma(i)}$.
If our variables are $x_{i,j}$ for a permutation matrix, then the cost of a route would be $\sum_{i,j,t} d_{i,j} x_{i,t} x_{j,t\oplus 1}$.
Fixing the first node in the route will reduce redundancy in the description by removing cyclical solutions from the search space.
To ensure our variables truly form a permutation matrix, we must then also have the constraints $\forall i\, \sum_j x_{i,j} = 1$ and $\forall j\, \sum_i x_{i,j}=1$.
At this stage, we could reduce the MILP formulation into a QUBO using previously discussed techniques. This would then increase the search space of the solutions from $n! = \mathcal{O} ( 2^{n\log(n)})$ to $2^{n^2}$. Instead we propose modifying existing algorithms from \cref{quantum algorithms} to preserve the structure of permutation matrices such that for all possible circuit parameters, $U(\vec{\theta}) \ket{\psi_0}$ only has support on computational states which are permutation matrices.

\subsection{Motivation}

The subspace of the Hilbert Space where our constraints are satisfied is spanned by computational states which are also permutation matrices, $\mathcal{P} = \{\ket{\psi} | \ket{\psi} = \sum_{\sigma\in P} \alpha_\sigma \ket{\sigma}\}$. If we choose to build $U(\vec{\theta})$ of gates that leave $\mathcal{P}$ as an invariant subspace, then $U(\vec{\theta})\ket{\psi_0} \in \mathcal{P} \iff \ket{\psi_0} \in \mathcal{P}$. So it would suffice to prepare the initial state to be a superposition of permutation matrices, such as $\ket{\psi_0} = \frac{1}{\sqrt{n!}} \sum_{\sigma\in P} \ket{\sigma}$, which can be achieved efficiently \cite{berry2018improved}. 
In our VQE circuit and QAOA inspired circuit, we carefully choose the gates in $U(\vec{\theta})$ to have this invariance property \cite{fuchs2022constraint}. Since our desired subspace contains a lot of symmetry, there are many natural sets of gates that have the required property. As the computational states are permutation matrices, any operation that preserves permutation matrices will have this property. A natural choice here would be permuting the rows and columns. Let $(\sigma, \tau)$ be a pair of permutation matrices. Then the unitary defined by $V_{(\sigma, \tau)}: \ket{p}\rightarrow \ket{\sigma p \tau}$ would leave $\mathcal{P}$ invariant. Moreover, this gate would be comprised only of performing swap gates on the qubits and is efficiently implementable.
In order to have parameterisable gates, we restrict ourselves to the case where $\sigma^2 =\tau^2 =\mathbb{I}$. This  way we have $V_{(\sigma, \tau)}^2 =\mathbb{I}$, and we can use Quantum Phase Estimation \cite{d1998general} to implement $e^{i\theta V_{(\sigma,\tau)}}$.
The familiar instances where $V_{(\sigma, \tau)}^2=\mathbb{I}$ would be when a single pair of rows or columns of the permutation matrix are swapped.

\subsection{Q-SWAP}

We take inspiration from this formalism and make changes that should allow us to optimise entirely within the space of feasible solutions and also allow for more adaptability in the search space, allowing us to arrive at the ground state faster. 

\begin{figure}[h]
\label{Q-SWAP Circuit}
\begin{quantikz}[row sep = 0.1cm]
\lstick[8]{$\ket{\psi_0} =\sum_{p\in P_n}\ket{p}$}& \gate[8]{e^{i \delta H}} & \gate[8]{e^{i \theta_0 V_{\sigma_0, \tau_0}}} & \ldots  \ldots \ldots
& \gate[8]{e^{i \delta H}} & \gate[8]{e^{i \theta_k V_{\sigma_k, \tau_k}}} & \gate[8]{e^{i \delta H}} & \gate[8]{e^{i \theta V_{\sigma, \tau}}} &\meter{} \rstick[8]{$\,\,\,\,\,\,\,\,\,\,\, \langle H \rangle $}\\
&&& \ldots  \ldots \ldots &&&&& \meter{}\\
&&& \ldots  \ldots \ldots &&&&& \meter{}\\
&&& \ldots  \ldots \ldots &&&&& \meter{}\\
&&& \ldots  \ldots \ldots &&&&& \meter{}\\
&&& \ldots  \ldots \ldots &&&&& \meter{}\\
&&& \ldots  \ldots \ldots &&&&& \meter{}\\
&&& \ldots  \ldots \ldots&&&&&  \meter{}\\
\end{quantikz}
\caption{Q-SWAP circuit. A classical optimiser is used to tune the optimal $\theta, \sigma,\tau$. Once this is done, they are added to the circuit and a new layer $e^{i \delta H},\, e^{i \theta V_{\sigma, \tau}}$ is added to optimise over.}
\label{Q-SWAP}
\end{figure}
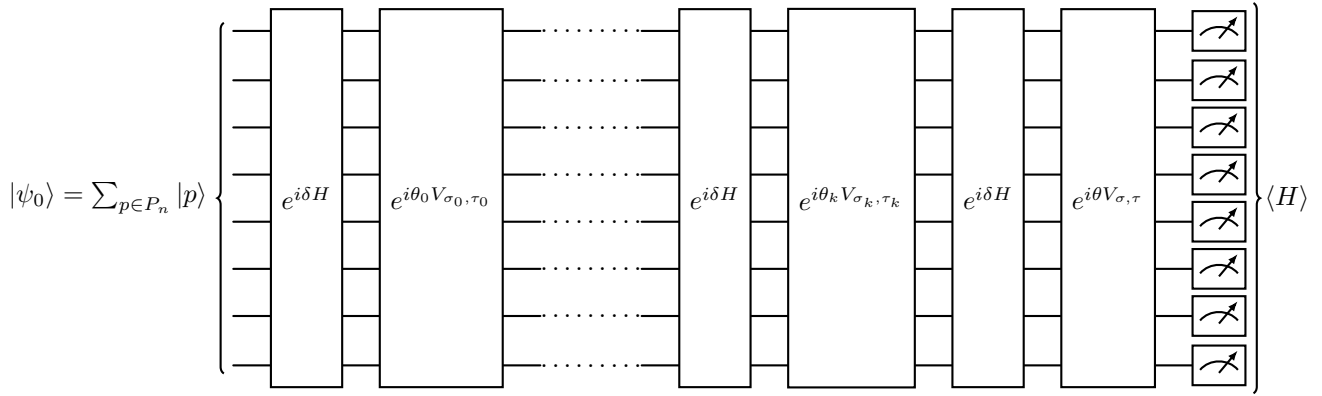

The algorithm Q-SWAP, shown in \cref{Q-SWAP},  is defined as follows:

\begin{enumerate}
    \item Prepare the initial state $\ket{\psi}:=\ket{\psi_0} = \frac{1}{n!} \sum_{p\in P} \ket{p}$ and set $counter= 0$
    \item Apply the gate $e^{i\delta H}$
    \item Obtain a pair of permutations $\sigma, \tau$ (can be random or through an optimisation process)
    \item Estimate $f(\theta) = \bra{\psi} e^{-i \theta V_{(\sigma, \tau)}} H e^{i \theta V_{(\sigma, \tau)}} \ket{\psi}$ for $\theta =  0, \frac{\pi}{4}, \frac{\pi}{2}$
    \item Find values of $A, B, \phi $ such that $f(\theta) = A+ B\cos(2\theta - \phi)$ and compute $\theta^* = \frac{\phi + \pi}{2}$
    \item Apply $e^{i\theta^* V_{\sigma, \tau}}$ to $\ket{\psi}$
    \item $counter \rightarrow counter + 1$, if $counter \leq steps$, return to 2
    \item Take multiple measurements of $\ket{\psi}$ and output the route with the minimal cost.
    
\end{enumerate}
Remarks:
\begin{enumerate}
\item 
The justification for step 4, is that since $V^2 = \mathbb{I}$, we can decompose $\ket{\phi} = \alpha \ket{V_+} + \beta \ket{V_-}$. The computation then becomes:
\begin{align*}\bra{\psi} e^{-i \theta V} H e^{i \theta V} \ket{\psi} &= (e^{i\theta V}(\alpha \ket{V_+} + \beta \ket{V_- }))^\dagger H (e^{i\theta V}(\alpha \ket{V_+} + \beta \ket{V_- })\\ 
&= (e^{i\theta }\alpha \ket{V_+} +  e^{-i\theta}\beta \ket{V_- }))^\dagger H (e^{i\theta}\alpha \ket{V_+} + e^{-i\theta} \beta \ket{V_- })\\
&= |\alpha|^2 \bra{V_+} H \ket{V_+} + |\beta|^2 \bra{V_-} H \ket{V_-} +  e^{2i\theta} \alpha \beta^* \bra{V_-}H\ket{V_+} + e^{-2i\theta} \alpha^* \beta \bra{V_+}H\ket{V_-}\\
&= |\alpha|^2 \bra{V_+} H \ket{V_+} + |\beta|^2 \bra{V_-} H \ket{V_-} + 2\mathfrak{R}(e^{2i\theta} \alpha \beta^* \bra{V_-}H\ket{V_+})\\
&= A + B \cos(2\theta - \phi)
\end{align*}
where $A =|\alpha|^2 \bra{V_+} H \ket{V_+} + |\beta|^2 \bra{V_-} H \ket{V_-} $ and $Be^{-i\phi} = 2 \alpha \beta^* \bra{V_-}H\ket{V_+})$.

Other values of $\theta$ can be sampled to increase the convergence of $\phi$, which is the only parameter we require for optimisation, but $\theta = 0, \frac{\pi}{4}, \frac{\pi}{2}$ will give us estimates for $A+ B\cos(\phi), A+ B \sin(\phi), A - B\cos(\phi)$. which can be used to calculate $A, B, \phi$ directly.
\item 
Optimising $\sigma, \tau$ to yield the largest decrease in expectation can be done using classical algorithms and taking expectations. The advantage to this compared to randomly picking permutations is that this will allow the circuit to have a lower depth, which is important to keep in mind for NISQ devices.
The topic of strategies for choosing optimal $\sigma, \tau$ is unclear and the focus of future works. The simple strategies employed by us were:
\begin{itemize}
    \item Random 1 swap - Here we fix $\tau = \mathbb{I}$ and aim to pick a strong $\sigma$. We try this by uniformly sampling from permutations of order 2, many times and then choosing $sigma$ with the largest drop in expectation.
    \item Random both sides -  Choose a random pair of initial permutations $\sigma_0, \tau_0$. We take it in turns optimizing over $\sigma$, then $\tau$, we randomly sample distance 2 permutations and if $V_{\sigma, \tau_0}$ yields a greater decrease in expectation, we update $\sigma_0 \rightarrow \sigma$. We do the same process for $\tau_0$ and repeat for a fixed number of rounds. 
    \item Mutations - Choose a random pair of initial permutations $\sigma_0, \tau_0$. Consider the pairs of permutations of distance 1 from them and update $\sigma_0$ and $\tau_0$, if they cause a larger drop in expectation. we reapeat this process with our new pair $(\sigma, \tau)$, until we reach a local minima.
\end{itemize}
\item $\delta$ is chosen to be small enough to scale the eigenvalues of $H$ between $0$ and $\pi$. This is an attempt to discourage  destructive interference between solutions with low energy eigenvalues, and maximize constructive interference for states with high energy eigenvalues.
\end{enumerate}

\subsection{Variational Quantum Eigensolver}

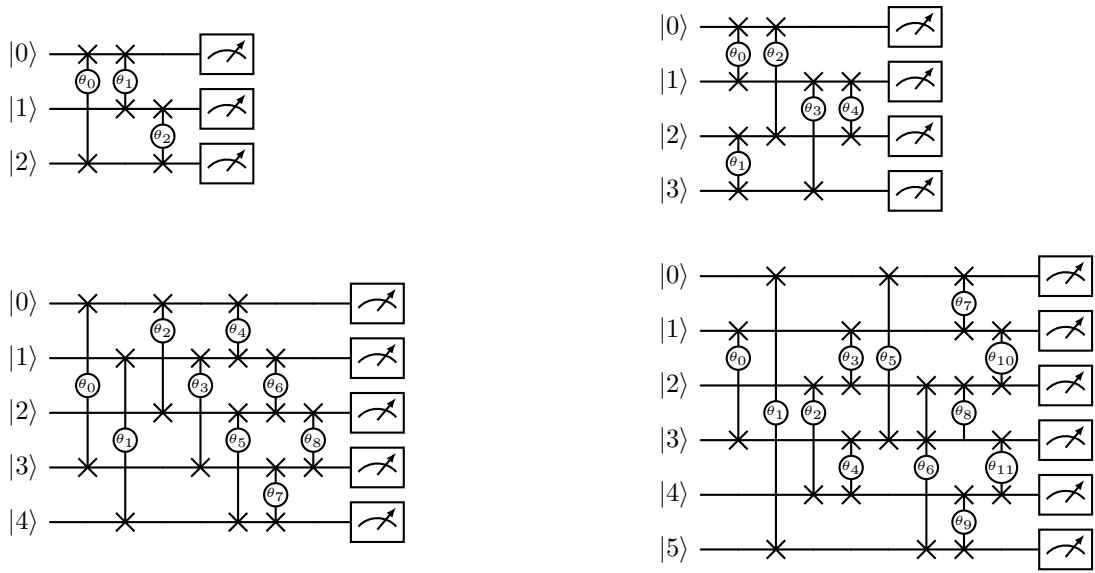
\begin{figure}[ht] \label{ fig7} 
  \begin{minipage}[b]{0.5\linewidth}
\begin{quantikz}[row sep=0.2cm]
\lstick{$\ket{0}$} &\swap[partial swap={\theta_0},partial position=0.25]{2}  &\swap[partial swap={\theta_1},partial position=0.5]{1}  & & \meter{}\\
\lstick{$\ket{1}$} && \targX{}  &\swap[partial swap={\theta_2},partial position=0.5]{1}&\meter{} \\
\lstick{$\ket{2}$} & \targX{}  &&\targX{} & \meter{}\\
\end{quantikz}
  \end{minipage} 
  \begin{minipage}[b]{0.5\linewidth}

\begin{quantikz}[row sep=0.2cm]
\lstick{$\ket{0}$} &\swap[partial swap={\theta_0},partial position=0.5]{1} &\swap[partial swap={\theta_2},partial position=0.25]{2} &&& \meter{} \\
\lstick{$\ket{1}$}& \targX{}  & & \swap[partial swap={\theta_3},partial position=0.25]{2} &  \swap[partial swap={\theta_4},partial position=0.5]{1}  & \meter{}\\
\lstick{$\ket{2}$}&\swap[partial swap={\theta_1},partial position=0.5]{1} & \targX{}  && \targX{}  &\meter{}\\
\lstick{$\ket{3}$}&\targX{}  && \targX{}  && \meter{}\\
\end{quantikz}
  \end{minipage} 
  \begin{minipage}[b]{0.5\linewidth}
  
\begin{quantikz}[row sep=0.2cm]
\lstick{$\ket{0}$} &\swap[partial swap={\theta_0},partial position=0.5]{3}&& \swap[partial swap={\theta_2},partial position=0.25]{2}&& \swap[partial swap={\theta_4},partial position=0.5]{1} &&& \meter{} \\
\lstick{$\ket{1}$} &&\swap[partial swap={\theta_1},partial position=0.5]{3}&&\swap[partial swap={\theta_3},partial position=0.25]{2}&\targX{}&  \swap[partial swap={\theta_6},partial position=0.5]{1} &&\meter{}\\
\lstick{$\ket{2}$} &&&\targX{}  &&  \swap[partial swap={\theta_5},partial position=0.25]{2}& \targX{}  & \swap[partial swap={\theta_8},partial position=0.5]{1} &\meter{} \\
\lstick{$\ket{3}$} &\targX{}  &&&\targX{} &&  \swap[partial swap={\theta_7},partial position=0.5]{1} & \targX{}  &\meter{} \\
\lstick{$\ket{4}$} &&\targX{}  &&& \targX{}  &  \targX{}  && \meter{} \\
\end{quantikz}
  \end{minipage}
  \hfill
  \begin{minipage}[b]{0.5\linewidth}
  
\begin{quantikz}[row sep=0.2cm]
\lstick{$\ket{0}$} &&\swap[partial swap={\theta_1},partial position=0.5]{5} &&& \swap[partial swap={\theta_5},partial position=0.5]{3}&&\swap[partial swap={\theta_7},partial position=0.5]{1}&&\meter{} \\
\lstick{$\ket{1}$} &\swap[partial swap={\theta_0},partial position=0.25]{2}&&&\swap[partial swap={\theta_3},partial position=0.5]{1}&&& \targX{}  &\swap[partial swap={\theta_{10}},partial position=0.5]{1}&\meter{}\\
\lstick{$\ket{2}$} &&&\swap[partial swap={\theta_2},partial position=0.25]{2}&\targX{}  && \swap[partial swap={\theta_6},partial position=0.5]{3}&\swap[partial swap={\theta_8},partial position=0.5]{1}& \targX{}  &\meter{}\\
\lstick{$\ket{3}$} &\targX{}  &&& \swap[partial swap={\theta_4},partial position=0.5]{1}& \targX{}  &\targX{}  && \swap[partial swap={\theta_{11}},partial position=0.5]{1}&\meter{}\\
\lstick{$\ket{4}$} &&&\targX{}  & \targX{}  &&&\swap[partial swap={\theta_9},partial position=0.5]{1}& \targX{}  & \meter{}\\
\lstick{$\ket{5}$} &&\targX{}   &&&& \targX{}  & \targX{}  &&\meter{}\\
\end{quantikz}

  \end{minipage} 
  \caption{Circuit ansatzes for a various number nodes that can achieve all possible groundstates for the time indexed hamiltonian formulation of the travelling salesman problem. Parameters start randomised and then trained to reduce the expected cost of the route. }
  \label{sorting-networks}
\end{figure}

Here we demonstrate a variable-efficient VQE ansatz to allow us to search only the state space of valid journeys. The circuit is achieved by initialising $n$ registers of length $n$ in a valid permutation, i.e. $\ket{0}\ket{1}\ket{2}...\ket{n-1}$. Our ansatz circuit is chosen as a combination of parameterised swap gates, which will then allow us to reach all other possible permutations. To achieve this with minimal gates/parameters, we have chosen to use minimal sorting networks\cite{batcher1968sorting} (the minimal number of swaps needed to reach all possible permutations) as shown in \cref{sorting-networks}. The number of parameters needed scale is $\mathcal{O}(n\log(n))$, and again we have the advantage that our state is constrained to always have support on valid states and no additional constraints are needed in the cost function. Here, for simplicity, to calculate the expectation $\bra{\psi}H\ket{\psi}$, we perform many measurements of $\ket{\psi}$ to sample from states distribution distribution $\{x: |\bra{x} \ket{\psi}\|^2 , x \in \mathcal{P}_n \}$ which can the be used to compute an estimate for expected cost of the route.
There is an interesting trade off between this and the Q-SWAP approach. Here, the number of parameters in the model are fixed at the start circuit and then optimised together. In Q-SWAP, however, the number of variables is linear in the depth of the circuit, which depends on how well the solution can be found. In Q-SWAP, each variable is also individually maximized, which is not the case here.

\subsection{Results}
We simulated Q-SWAP and VQE approaches with a noiseless simulator to solve travelling salesman instances on six and seven nodes as in \cref{Toy TSP}. The results for how the AR varied over the computation is shown in \cref{AR-TSP}. There are two main takeaways from the results. 
Firstly, there is a clear advantage to putting care into how $\sigma, \tau$ are chosen, which allowed the mutations style approach to vastly outperform sampling random permutations. It is still not clear what would be the best approach to choosing optimal permutations to perturb the state by, but this will studied further in future work.
Secondly, the plateau of the AR will be correlated to number of nodes in the instance. Each iteration of $e^{iV_{\sigma, \tau} \theta}$  is simultaneously moving more amplitude into low energy states and moving amplitude from low energy states to higher ones. When a certain threshold of amplitude is congregated into low energy states, it would be beneficial to not change the amplitudes and to choose $\theta =0$. When this threshold has occurred, we must already have a certain amount of amplitude in good states so we can, in principle, have a non-trivial percentage of measuring close to optimal routes. How these features are correlated merits further research to fully understand.
The results for VQE approach in \cref{AR-TSP-VQE}, are very promising with the solver converging perfectly to the ground state for the case of six nodes. In both cases there clear points when the solution begins to plateau before further decreasing. This is when the parameters tune towards a certain which is a local minimum, which the classical optimiser then took many steps to escape. Using different sorting networks with possible degeneracy could potentially create a route to escaping these barren plateaus, and will be further investigated \cite{uvarov2021barren}.

\begin{figure}[h]

\begin{subfigure}{0.5\textwidth}

\begin{tikzpicture}[
    ->, 
    >=Stealth, 
    every node/.style={draw,minimum size=0.2cm, font=\sffamily\Large\bfseries} 
,scale = 0.2]
\small 
\node[font=\tiny] (BASE) at (0, 0) {B};
\node[font=\tiny] (0) at (-10.558076410502618, -15.873358630771367) {0};
\node[font=\tiny] (1) at (-4.1576702955727605, -13.80110916790359) {1};
\node[font=\tiny] (2) at   (-17.339396172816404, -3.9363594205970074) {2};
\node[font=\tiny] (3) at  (16.718201723508756, 12.018094059832343) {3};
\node[font=\tiny] (4) at  (10.606504100217538, -11.122872972387295) {4};

\draw (BASE) -- (2);
\draw (2) -- (0);
\draw (0) -- (1);
\draw (1) -- (4);
\draw (4) -- (3);
\draw (3) -- (BASE);

\end{tikzpicture}
\caption{TSP instance with 6 nodes}
\label{fig:subim1}
\end{subfigure}
\begin{subfigure}{0.5\textwidth}

\begin{tikzpicture}[
    ->, 
    >=Stealth, 
    every node/.style={draw,minimum size=0.2cm, font=\sffamily\Large\bfseries} 
,scale = 0.2]
\small 
\node[font=\tiny] (BASE) at (0, 0) {B};
\node[font=\tiny] (0) at (7.280182423519115, -16.335895676817444) {0};
\node[font=\tiny] (1) at (4.712653954456094, 13.676796182038245) {1};
\node[font=\tiny] (2) at  (13.382011543043589, 0.6007090316539774) {2};
\node[font=\tiny] (3) at  (5.241518611827061, -5.2308066374832585) {3};
\node[font=\tiny] (4) at (1.1207448807689886, -15.685732667890724) {4};
\node[font=\tiny] (5) at (7.317997353331862, 4.026734469991197) {5};

\draw (BASE) -- (1);
\draw (1) -- (5);
\draw (5) -- (2);
\draw (2) -- (0);
\draw (0) -- (4);
\draw (4) -- (3);
\draw (3) -- (BASE);

\end{tikzpicture}
\caption{TSP instance with 7 nodes}
\label{fig:subim2}
\end{subfigure}

\caption{TSP instances of sizes 6 and 7 nodes. Journeys start and end at the base, B.}
\label{Toy TSP}
\end{figure}

\begin{figure}[h]

\begin{subfigure}{0.5\textwidth}
\includegraphics[scale=0.75]{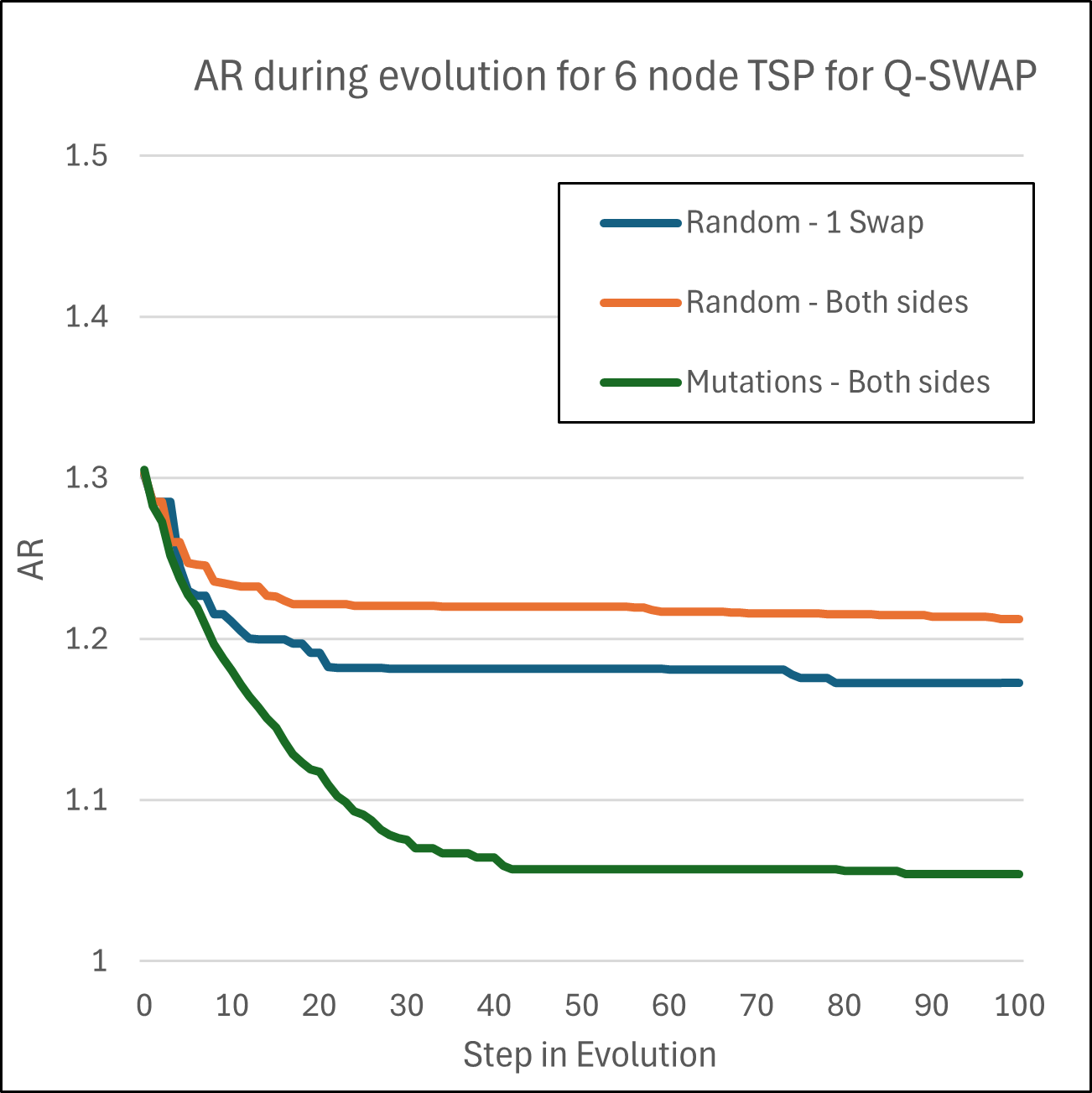}

\label{fig:subim1}
\end{subfigure}
\begin{subfigure}{0.5\textwidth}
\includegraphics[scale=0.75]{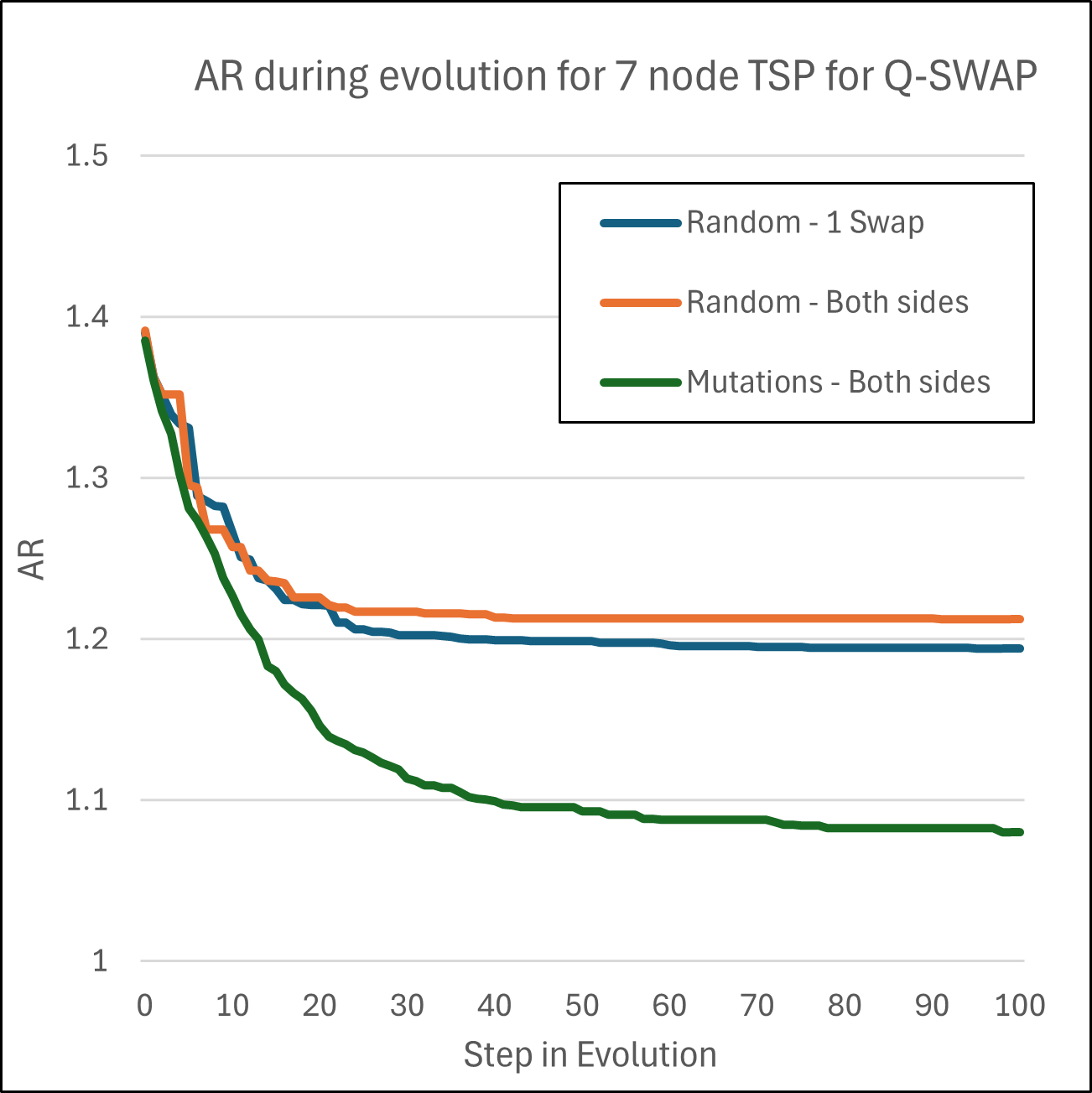}
\label{fig:subim2}
\end{subfigure}

\caption{Results from solving TSP instances from \cref{Toy TSP} using Q-SWAP. Here, three different approaches are used to determine how $\sigma, \tau$ are chosen for each stage. }
\label{AR-TSP}
\end{figure}

\begin{figure}[h]

\begin{subfigure}{0.5\textwidth}
\includegraphics[scale=0.75]{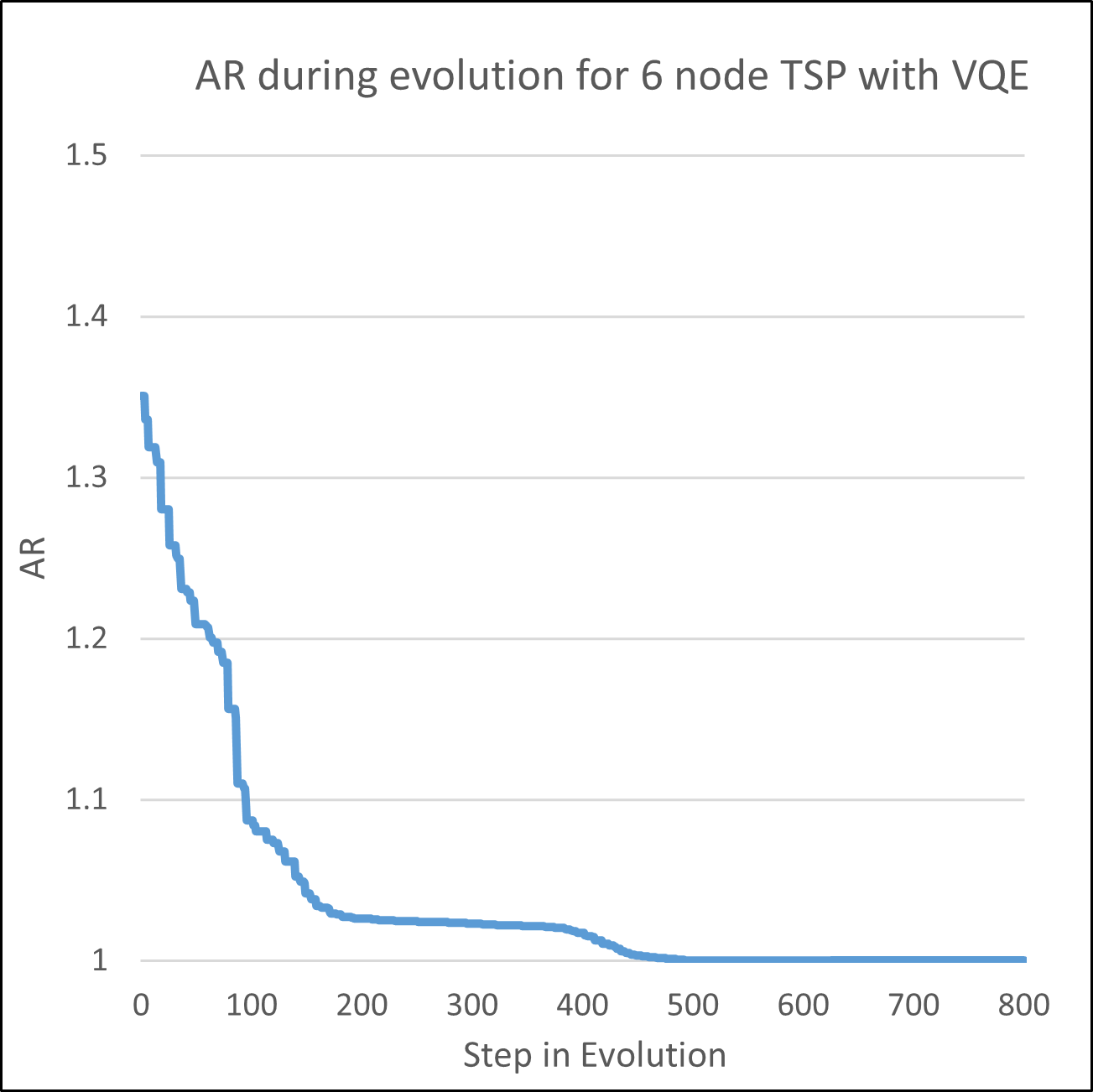}

\label{fig:subim1}
\end{subfigure}
\begin{subfigure}{0.5\textwidth}
\includegraphics[scale=0.75]{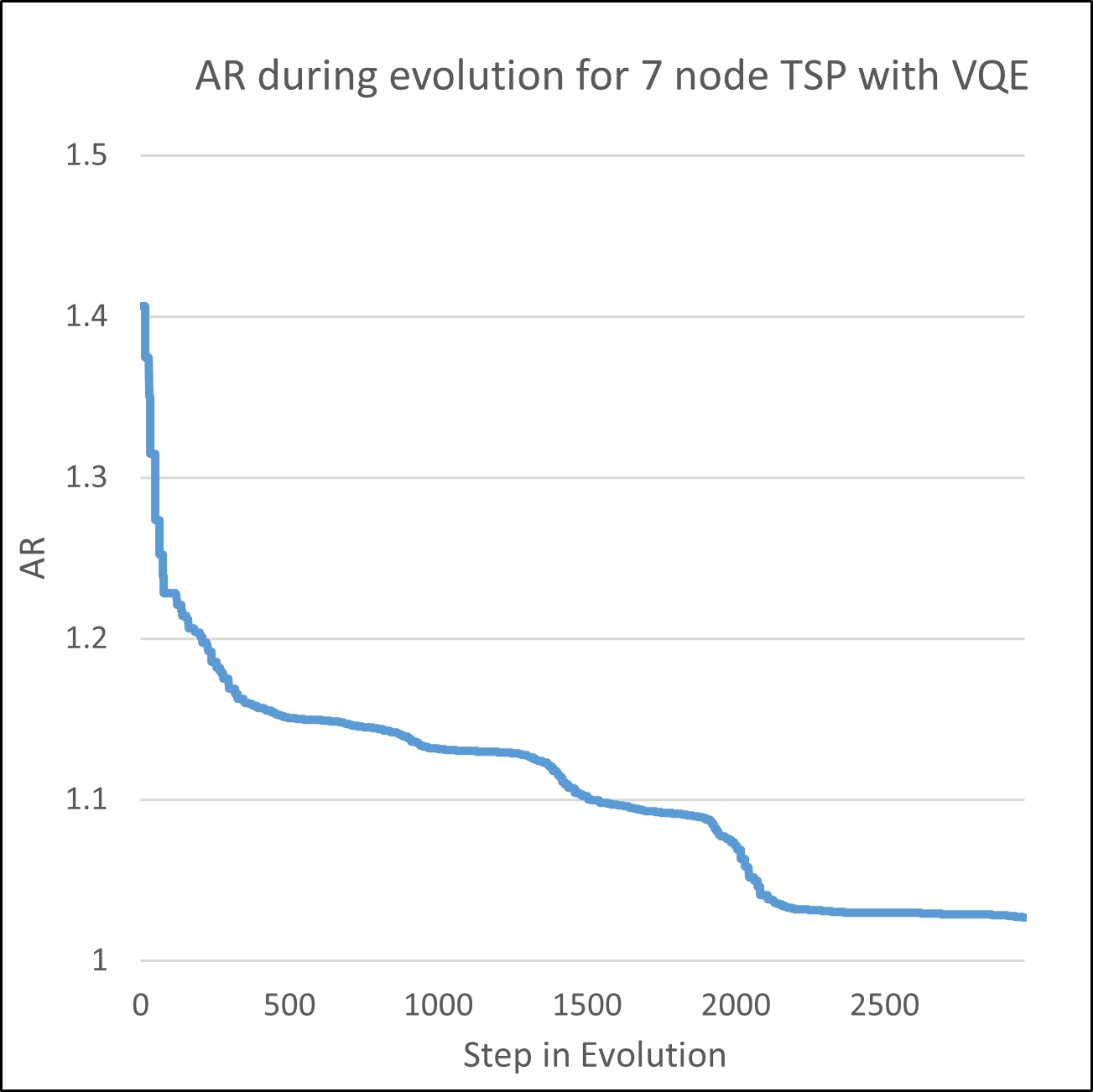}
\label{fig:subim2}
\end{subfigure}

\caption{Results from solving TSP instances from \cref{Toy TSP} using VQE with sorting network ansatz.}
\label{AR-TSP-VQE}
\end{figure}

\section{conclusion}

We have demonstrated a new MILP formulation for multiple drone routing problems which is flexible to accommodate for a plethora of interesting constraints which weren't simultaneously met by existing models. We have showed that this model can be used to find optimal solutions using current classical solvers and have showed the potential for quantum annealing to offer a speedup for this problem in the future. We have also shown the merit in customising standard quantum algorithms towards naturally meeting constraints, leading to a reduction in the size of the search space and number of parameters to optimise over.
\newline

\bibliography{main_SPIE} 
\bibliographystyle{spiebib} 

\end{document}